\documentclass[aps,showpacs,prl,floatfix,twocolumn,amsmath,citeautoscript,longbibliography]{revtex4-1}
\usepackage{graphicx}
\usepackage{amsmath}
\usepackage{amssymb}
\usepackage{bm}
\usepackage{dcolumn}
\usepackage{subfigure}
\usepackage{units}
\usepackage{hyperref}
\usepackage[usenames]{color}
\usepackage{booktabs}
\usepackage{multirow}
\usepackage{siunitx}
\usepackage{wrapfig}
\hypersetup{pdfborder=0 0 0,colorlinks=true,citecolor=blue,linkcolor=blue}
\usepackage{natbib}

\newcommand{\MX}{MX$_2$}
\newcommand{\MS}{MoS$_2$}

\newcommand{\MSe}{MoSe$_2$}

\begin{document}

\title{The Sign of Three: Spin/Charge Density Waves \\ at the  Boundaries of Transition Metal Dichalcogenides}

\author{Sridevi Krishnamurthi and Geert Brocks}%
\email[]{g.h.l.a.brocks@utwente.nl}
\affiliation{Computational Materials Science, Faculty of Science and Technology and MESA+ Institute for Nanotechnology, University of Twente, the Netherlands. }
\date{\today}
\begin{abstract}
One-dimensional grain boundaries of two-dimensional semiconducting {\MX} (M= Mo,W; X=S,Se) transition metal di-chalcogenides are typically metallic at room temperature. The metallicity has its origin in the lattice polarization, which for these lattices with $D_{3h}$ symmetry is a topological invariant, and leads to one-dimenional boundary states inside the band gap. For boundaries perpendicular to the polarization direction, these states are necessarily 1/3 occupied by electrons or holes, making them susceptible to a metal-insulator transition that triples the translation period. Using density-functional-theory calculations we demonstrate the emergence of combined one-dimensional spin density/charge density waves of that period at the boundary, opening up a small band gap of $\sim 0.1$ eV. This unique electronic structure allows for soliton excitations at the boundary that carry a fractional charge of $\pm 1/3\ e$.
\end{abstract}
\maketitle

{\em Introduction.} The two-dimensional transition metal di-chalcogenides (TMDCs) {\MX} (M= Mo,W; X=S,Se,Te), in their common $H$-structure, are semiconductors with band gaps of 1-2 eV. Surprisingly, many edges and grain boundaries of these TMDCs are metallic at room temperature \cite{PhysRevLett.87.196803}. This seems to be true irrespective of the substrate on which the TMDC is deposited, or whether UHV conditions are used or not \cite{Ma,Barja:2016aa}. Indeed, the exact atomic termination of edges and boundaries seems to be irrelevant for their metallicity \cite{PhysRevB.67.085410}. 

\begin{figure}[ht]
\includegraphics[width=8.5cm]{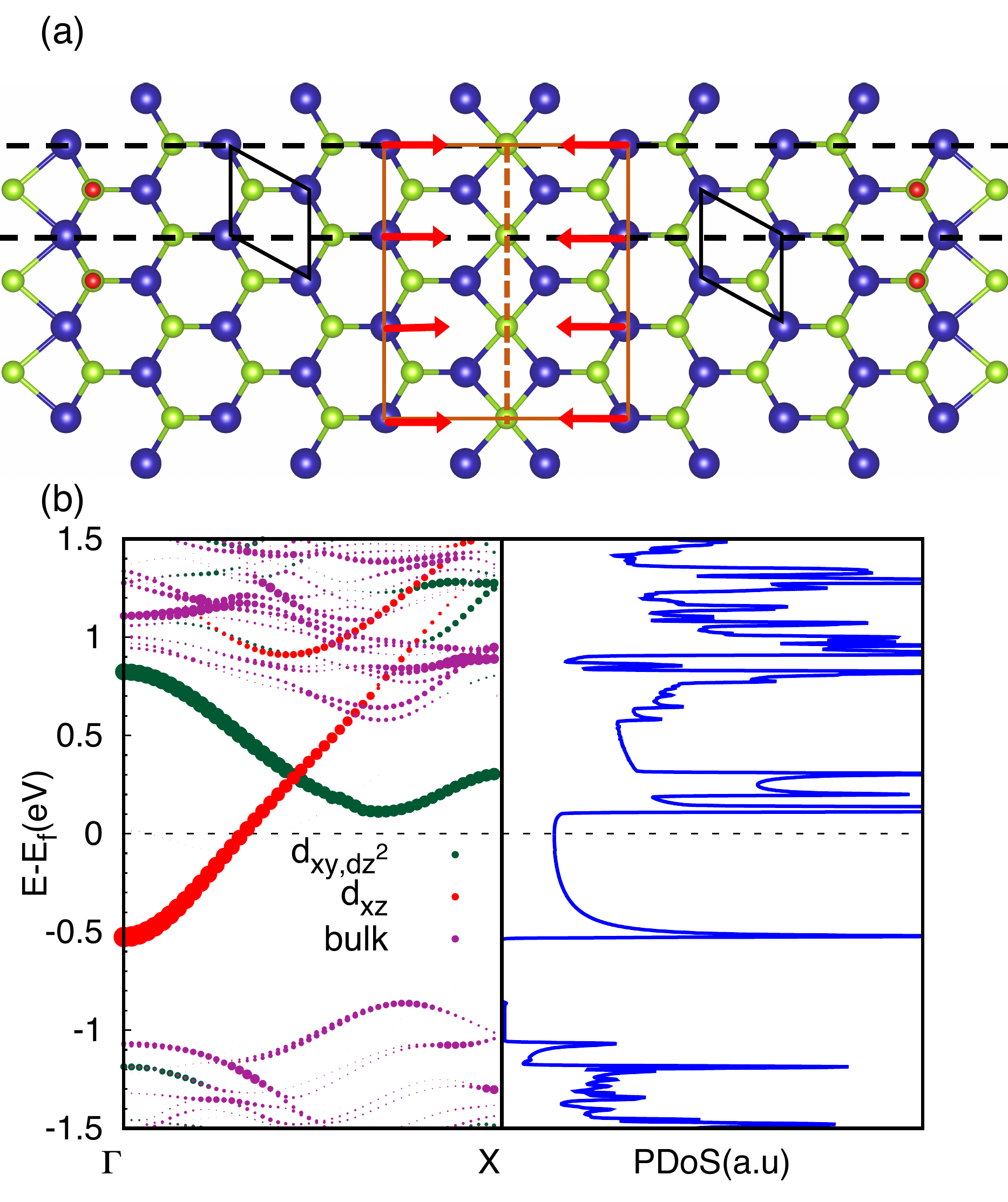}
\caption{(a) The 4$|$4P {\MX} MTB structure; the red and black dashed lines indicate the position of the MTB, and the periodicity along the MTB, respectively. The arrows indicate the polarization $\mathbf{P}$ direction, which inverts across the MTB. The blue, green, and red spheres indicate Mo, Se, and O atoms, see Supporting Information. (b) Calculated dispersion of DFT bands of {\MSe} along the MTB direction. The two bands in the band gap (red and green), originate from the Mo atoms at the MTB, and have $d_{xz}$ and $d_{xy}$/$d_{z^{2}}$ character. (c) The calculated density of states (DoS), with the MTB bands and their typical 1D van Hove singularities.}
\label{figure1}
\end{figure}

Experimentally, of the different possible TMDC edge configurations possible, mirror twin boundaries (MTBs) have been studied  most extensively \cite{Batzill_2018,Ma, Barja:2016aa, JamesC}. MTBs occur spontaneously when TMDC monolayers are grown on isotropic substrates, or on substrates with a high in-plane symmetry, such as graphite. In essence, a MTB is formed between two TMDC crystallites that have their crystal growth directions rotated by 60 degrees, thereby forming mirror images of one another along the line of coalescence, see Fig.~\ref{figure1}. MTBs are among the most predominant one-dimensional (1D) defects occurring during TMDC growth. 

Whereas the presence of MTBs can be desirable or undesirable from the point of view of applications, the 1D metallic nature of the MTBs makes them interesting from a fundamental perspective. Two distinct views exist on the basic electronic structure of such MTBs. From angle-resolved photoemission (ARPES) results on monolayer {\MSe}, the existence of a  Tomonaga-L\"{u}ttinger liquid (TLL) has been put forward \cite{ncomms-ma}, which has also been claimed from scanning tunnelling microscopy and spectroscopy (STM, STS) on finite-length MTBs \cite{PhysRevX.9.011055}. In contrast, low-temperature STM and STS on the same material demonstrate the presence of a charge density wave (CDW) at the MTB with a wavelength of three lattice constants,  opening up a band gap of $\sim 0.1$ eV \cite{ncomms-ma,Barja:2016aa}.

Density-functional-theory (DFT) first-principles studies of MTBs have focused foremost on their atomic structure, and their stability and formation energies \cite{Krasheninnikov}. There are several possible MTB structures, but overall the stoichiometric 4$|$4P structure, shown in Fig. \ref{figure1}, seems to occur most often experimentally \cite{komsa-MTB,Barja:2016aa}. DFT calculations predict this structure to be metallic if a periodicity of one lattice constant along the MTB is assumed. Although it may appear likely that such 1D metallic structures are susceptible to CDW Peierls distortions \cite{Barja:2016aa}, so far, first-principles calculations have not been able to identify the presence of such structural distortions at MTBs, without resorting to artificial displacements of atoms. 

In the present work, we analyze the electronic structure of 4$|$4P MTBs in {\MSe} and {\MS} monolayers. The metallicity is carried by 1D states localized at the MTB, where the intrinsic electric polarization of the 2D TMDC dictates a total occupancy of these states of $1/3$ per MTB lattice site. Including the spin degree of freedom, we show that a combined spin density wave (SDW) and charge density wave (CDW) at the MTB leads to a period tripling without structural distortion. The SDW/CDW lowers the total energy, and creates a band gap of a size comparable to experiment. The general mechanism proposed here not only holds for MTBs in {\MX} (M= Mo,W; X=S,Se,Te), but also for edges with zigzag orientations \cite{Lucking}, which are also commonly found in these materials. We speculate that this unique electronic structure allows for soliton excitations at such boundaries and edges that carry a fractional charge of $\pm 1/3\ e$ \cite{PhysRevB.85.073401}.

In our calculations we model  MTBs in a periodic supercell geometry. The supercell typically contains a ribbon of twelve {\MX} units across the $y$-direction and a number of units along the $x$-direction, with ribbons in neighbouring supercells separated by 10 \AA\ vacuum in the $y$ and $z$ direction, see Fig. \ref{figure1}. Further computational details can be found in the Supporting Information \footnote{ \href[page=2]{run:/supporting_information.pdf}{See Supporting Information} for details on exchange-correlation functionals used and other DFT parameters, It includes the references \cite{PhysRevB.54.11169,PhysRevB.59.1758,PhysRevB.50.17953,PhysRev.136.B864,PhysRev.140.A1133,PhysRevB.23.5048,PhysRevB.57.1505}}  \nocite{PhysRevB.54.11169,PhysRevB.59.1758,PhysRevB.50.17953,PhysRev.136.B864,PhysRev.140.A1133,PhysRevB.23.5048,PhysRevB.57.1505}.

{\em MTB states.} The {\MX} monolayer owing to its lack of inversion symmetry, has an in-plane electric polarization. Using the modern theory of polarization \cite{PhysRevB.47.1651,doi:10.1021/acs.nanolett.5b02834}, it has been shown that the polarization of lattices with $D_{3h}$ symmetry, such as the TMDCs discussed here, is a topological invariant \cite{PhysRevB.88.085110}. $D_{3h}$ symmetry only allows for polarizations  $\mathbf{P}=p_1\mathbf{a}_1+p_2\mathbf{a}_2$ with $(p_1,p_2)=(\alpha+n_1,\beta+n_2); n_{1,2}=0, \pm, 1,\pm 2 ... $, where $\mathbf{a}_{1,2}$ are the lattice vectors of the primitive 2D unit cell, and $(\alpha,\beta)$ is one out of three possible values: $(2/3,1/3)$, $(1/3,2/3)$, or $(0,0)$ (hence, the topological invariant is $\mathbb{Z}_3$). Straightforward DFT calculations show that all our {\MX} TMDCs belong to the same class and take on the specific value $(\alpha,\beta)=(2/3,1/3)$, see Fig. \ref{figure1}, which is in agreement with previous calculations \cite{doi:10.1021/acs.nanolett.5b02834}. 

Crossing the MTB,  the polarization is inverted ($\mathbf{P} \leftrightarrow -\mathbf{P}$), see Fig. \ref{figure1}. This abrupt jump in the topological invariant causes the semiconducting band gap to close, and gives rise to metallic states \cite{PhysRevB.99.155109}. These localised interface states are responsible for compensating the line charge ($\lambda = 2\mathbf{P}\cdot\hat{\mathbf{n}}$, with $\hat{\mathbf{n}}$ the normal to the MTB) that originates from the polarization \cite{PhysRevB.87.205423}. 
We stress that these are additional electronic gap states created as a result of the abrupt change of the topological invariant at the MTB, and not bulk bands pulled towards the Fermi level, as is sometimes argued to explain the metallic edges of a nanoribbon \cite{doi:10.1021/acs.nanolett.5b02834,PhysRevB.87.205423}, see Fig. \ref{figure1}. Indeed, similar states are found at edges modelled with a simple tight-binding model based on bulk parameters \cite{PhysRevB.93.205444,PhysRevB.99.155109}.
 
A DFT calculation of the band structure of a {\MSe} MTB with the smallest possible periodicity, shows two bands, one partially occupied and the other completely empty, that lie within the band gap, see Fig. \ref{figure1}, in agreement with previous calculations \cite{PhysRevX.9.011055}. MTBs of other {\MX} give similar band structures. On projecting on the atoms at the MTB, we find that one of these bands has mostly M $d_{xz}$ character (shown in red in Fig. \ref{figure1}), whereas the other band has mostly M $d_{xy}$/$d_{z^{2}}$ character (highlighted in green).

The occupancy of these MTB states can be deduced from a simple general reasoning. As the polarization jumps from $\mathbf{P}$ to $-\mathbf{P}$ going across the MTB, the result is a polarization line charge $\lambda = 2\mathbf{P}\cdot\hat{\mathbf{n}} = (2e)/(3a)$ at the MTB (where $a$ is the lattice constant along the MTB). In a system that consists of macroscopic domains separated by MTBs, all of these boundaries have to be neutral, such as to avoid a polarization catastrophe \cite{oxide}. This means that each MTB must also carry an electronic charge $-\lambda$, which compensates for the polarization charge. Such an electronic charge can only be carried by the 1D MTB interface states, located inside the band gap, as discussed above. 

This means that these particular bands must have a total occupancy of $2/3$ electrons. Referring to Fig. \ref{figure1}, this results in the lower of the two bands being $1/3$ filled (accounting for spin degeneracy), whereas the upper one is completely empty \footnote{Like in most DFT calculations, we model the MTB by a ribbon of a limited width. In such a geometry one can have a transfer of electrons between the (metallic) edges of the ribbon and the (metallic) MTB, in order to equilibrate the Fermi level. This can cause the occupancy of the MTB bands (and that of the edge bands) to deviate slightly from $1/3$, as can be observed in Fig. \ref{figure1}. Although this is physical for a small ribbon, the electron transfer is artificial when the ribbon is used to model a single MTB. We suppress this artificial electron transfer by forcing the edges to be insulating, which can be done in a supercell with tripled periodicity, \href[page=2]{run:/supporting_information.pdf}{See Supporting Information.}}.  Given their 1D character and the partial occupancy of $1/3$, these metallic states might then be susceptible to a Peierls-type structural distortion that leads to a tripling of the period. However, like previous calculations, our DFT calculations do not give such a spontaneous structural distortion of the MTB \cite{PhysRevX.9.011055,Barja:2016aa}.

{\em SDW/CDWs.} Nevertheless, it is highly unlikely that such 1D metallic states can escape electronic perturbations unscathed. We study the possibility of charge ordering and concomitant spin ordering using DFT+U calculations. The on-site electron-electron Coulomb interaction in 4d transition metal atoms (TMs), such as Mo, is supposed to be weaker than that in 3d TMs, and is thus often assumed to be negligible. Explicit calculations of the screened Hubbard $U$ in TMs \cite{PhysRevB.83.121101}, and in TM oxides \cite{PhysRevB.86.165105}, however show that the latter assumption is often not justified, and that a moderate value of $U\sim 3$ eV for Mo 4d states is not unreasonable. The states at the 1D grain boundaries have predominant Mo d character, which, because of the 2D surroundings, one may expect to be relatively weakly screened. It is therefore appropriate to include the on-site Coulomb interaction. 

\begin {figure}[htbp]
\includegraphics[width=9.0cm]{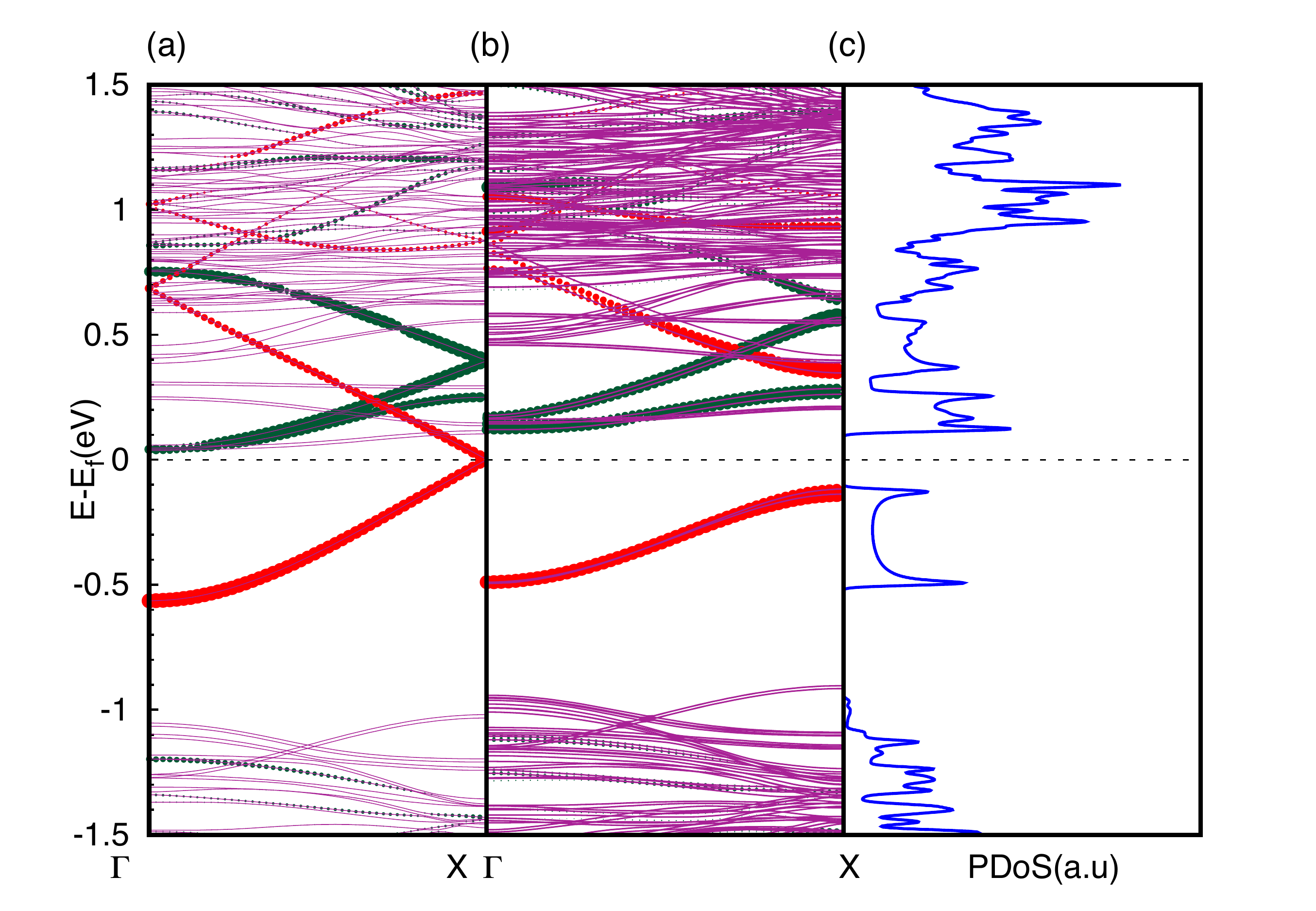}
\caption{ (a) Band structure of the undistorted MTB of {\MSe}, folded in the tripled cell. (b)  The SDW/CDW  opens up an indirect band gap of $\sim 0.26$ eV. (c) The corresponding DoS clearly shows this gap, and the typical 1D van Hove singularities. }
\label{figure2}
\end{figure}

As a starting point, Fig.~\ref{figure2}(a) shows the same {\MSe} band structure as in Fig.~\ref{figure1}, but now in a $3\times$ cell, i.e., a cell that is tripled along the direction of the MTB. The Mo $d_{xz}$ band (red) that was $1/3$ occupied in the simple unit cell, is now of course folded such, that the lowest branch is completely filled, and the upper two branches are completely empty. The Mo $d_{xy}/d_{z^{2}}$ band (green) is also folded twice, but its three branches lie above the Fermi level. Using a Hubbard $U-J=3$ eV \footnote{We use the rotationally invariant of the DFT+U functional, as formulated by Dudarev {\em et al.}, where the Hubbard $U$ and exchange $J$ are combined into one parameter $U-J$ \cite{Dudarev}.}, and re-optimizing the electronic structure \cite{PhysRevB.82.195128}, opens up a gap of $\sim 0.47$ eV between the filled and the empty states of the Mo $d_{xz}$ band, as can be observed clearly in Fig.~\ref{figure2}(b). The (empty) Mo $d_{xy}/d_{z^{2}}$ bands change very little. The result is a band structure showing an indirect gap of $\sim 0.26$ eV between the occupied Mo $d_{xz}$ band at X, and the unoccupied Mo $d_{xy}/d_{z^{2}}$ band at $\Gamma$. The corresponding DoS, shown in Fig.~\ref{figure2}(c), shows this band gap, clearly marked by van Hove singularities characteristic of 1D structures. The emergence of this SDW decreases the total energy of the MTB by $67$ meV/$3\times$ cell.

\begin {figure}[htbp]
\includegraphics[width=8.5cm]{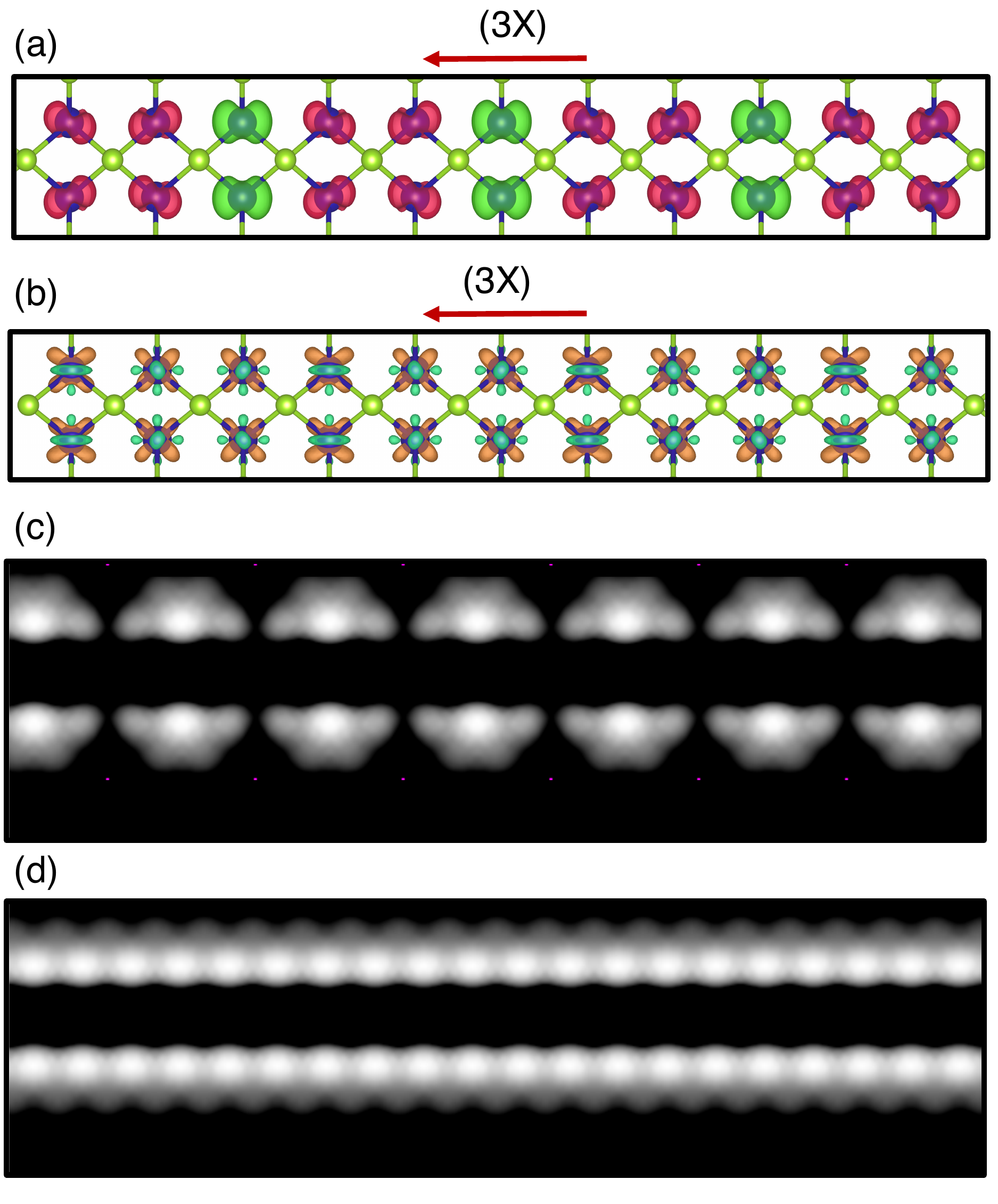}
\caption{ (a) Spin density at the MTB of {\MSe}; the red/green colors indicate spin up/down, where the SDW results in the three Mo atoms along the MTB becoming inequivalent\cite{vesta}. (b) The corresponding charge density; the brown/green colors indicate the change with respect to the charge density of the ideal $1\times$ structure. (c) LDoS, plotted as a simulated STM image \cite{PhysRevB.77.241308}, demonstrating the tripling of the translation period along the MTB. (d) For comparison, the LDoS of the ideal $1\times$ structure without the SDW/CDW is shown. Both LDoSs are integrated from $-0.5$ eV up to the Fermi level.}
\label{figure3}
\end{figure}

The origin of this band gap opening lies at the emergence of a SDW localized on the atoms closest to the MTB, which leads to magnetic moments on the three Mo atoms on one side of the MTB  of $0.40$, $-0.20$, and $-0.21$ $\mu_B$, respectively (the three Mo atoms on the other, the mirrored, side of the MTB have exactly the same magnetic moments). The inequivalence of the three Mo atoms is clearly visible in the spin density shown in Fig.~\ref{figure3}(a). This SDW is accompanied by a quite subtle CDW, as shown in the corresponding local density of states (LDoS) in Fig.~\ref{figure3}(b), which leads to a tripling of the period as observed in STM \cite{Barja:2016aa}, compare Fig.~\ref{figure3}(c). Although we let the geometry of the MTB free to relax with the SDW/CDW, we observe no visible distortion in the structure. For instance, the bond lengths and bond angles between all atoms along the MTB remain the same. This means that the SDW/CDW is a purely electronic effect, and not a Peierls distortion.

{\MS} behaves similarly to {\MSe}; in the 4$|$4P MTB structure of {\MS} a gap is opened by a SDW/CDW with $3\times$ periodicity. Using $U-J=3$ eV, the resulting magnetic moments on the Mo atoms closest to the MTB are $0.25$, $-0.21$, and $-0.05$ $\mu_B$. The moments are somewhat smaller than for {\MSe}, as is the induced gap at X. The resulting band structures in the gap region of {\MS} and {\MSe} are however quite similar, with {\MS} showing an overall indirect band gap of $0.10$ eV between the occupied Mo $d_{xz}$ band at X, and the unoccupied Mo $d_{xy}/d_{z^{2}}$ band at $\Gamma$. The total energy of the {\MS} MTB is decreased by $27$ meV per $3\times$ cell., \href[page=2]{run:/supporting_information.pdf}{see Supporting Information.}

\begin {figure}[htbp]
\includegraphics[width=8.0cm]{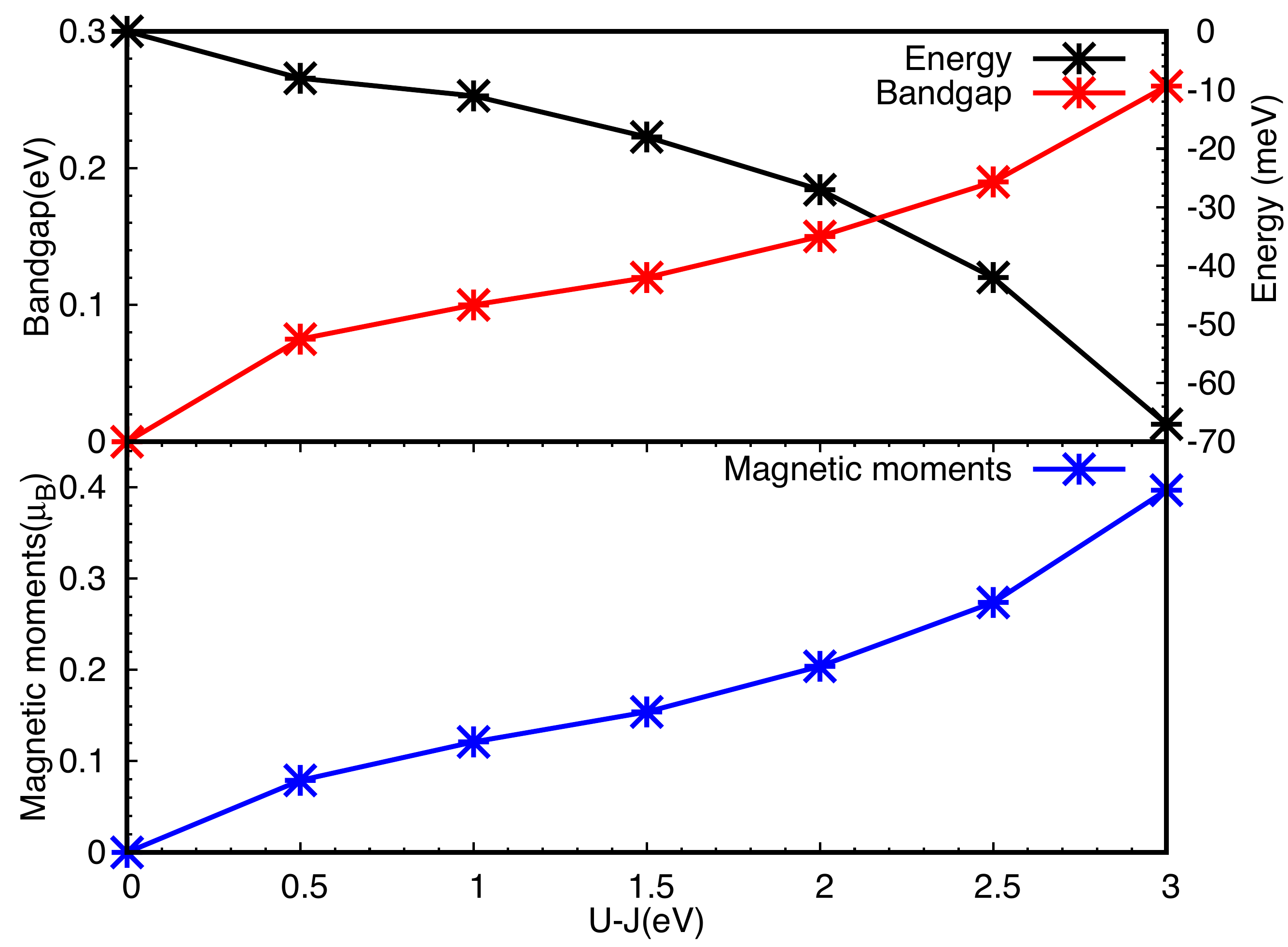}
\caption{ Top:  band gap (red) and total energy per $3\times$ cell (black) of the MTB of {\MSe}, as function of the Hubbard $U-J$ value. Bottom: maximal magnetic moment on the MTB Mo atoms as function of $U-J$.}
\label{figure4}
\end{figure}

The onsite electron-electron Coulomb interaction is essential for the development of a SDW/CDW, i.e., in a calculation with $U-J=0$ it does not happen. Figure \ref{figure4} shows the size of the band gap, the total energy decrease, as well as the size of the maximal magnetic moment on the Mo atoms at the MTB, as a function of the Hubbard $U-J$ value used in the calculation. It can be observed that both the band gap and the magnetic moments, increase monotonically with increasing $U-J$, whereas the total energy decreases monotonically. All, however, remain sizeable even for relatively small values of $U-J$, which indicates the robust presence of a SDW/CDW. Only if $U-J$ becomes smaller than $\sim 0.5$ eV, a SDW/CDW fails to develop.

Experimentally, the electronic structure of 4$|$4P {\MSe} MTBs has been interpreted using STM and STS in terms of CDWs by Barja {\em et al.} \cite{Barja:2016aa}, where they observed the characteristic $3\times$ periodicity. The observed band gap of $\sim 0.1$ eV suggest that the effective value of $U-J$ in their case is rather moderate, i.e., in the range 1-1.5 eV. Similarly, these CDWs have been seen in STM by Ma {\em et al.}, and characterized by means of temperature-dependent conductivity measurements. We suggest that the CDW is accompanied by a SDW, which, although the magnetic moments are moderate, may be observed using spin-polarized STM. 

Whereas the SDW/CDW should represent the ground state of the MTB, we cannot exclude that at a higher temperature, or for a markedly different MTB structure, electron correlations take over that are typical of 1D TLLs, as argued in Refs. \cite{ncomms-ma,PhysRevX.9.011055}.

In the mean time, SDW/CDWs of $3\times$ periodicity allow for interesting soliton excitations, i.e., localized quasi-particles with fractional charges $\pm 1/3\ e$ or $\pm 2/3\ e$, and spin $1/2$, $0$, or even an irrational number \cite{PhysRevLett.46.738,PhysRevLett.48.1416}. Such solitons will occur naturally on MTBs with an overall length that is not a mutiple of $3a$, because the boundary conditions at both ends of the MTB introduce frustration in the lattice \cite{PhysRevLett.109.246802}. In MTBs with lengths that are a multiple of $3a$, solitons do not exist in the ground state, but may be introduced by excitation. In particular, depositing TMDCs on substrates with which the electronic coupling is very weak, it may be possible to observe their fractional charges in a Coulomb blockade experiment, using STM, for instance.

In summary, using DFT+U calculations we have shown that a combined SDW and CDW of triple period arises in MTBs of TMDCs, which open up a band gap of $\sim 0.1$ eV in the 1D metallic band structure of a MTB. We argue that the triple period is necessarily the result of the topological invariant of these systems, i.e., the lattice polarization, which leads to metallic states in the 2D band gap, localized at the MTB, with a total occupancy of $1/3$. The emergence of a SDW/CDW lifts the metallicity, but it also allows for topological soliton excitations, with charges that are multiples of $1/3\ e$.

\bibliography{/Users/sridevi/Desktop/MTB-Paper/MTB_V3g/mtb_v3.bib}
 
\end{document}